\documentclass[preprint, authoryear, 1p]{elsarticle}

\usepackage{hyperref}
\usepackage{bm}
\usepackage{amsmath, amssymb}

\journal{arXiv}

\biboptions{authoryear}

\begin{document}

\begin{frontmatter}

\title{Selective inference after likelihood- or test-based model selection in linear models}

\author[aaa]{David R\"ugamer\corref{mycorrespondingauthor}}
\cortext[mycorrespondingauthor]{Corresponding author.}
\ead{david.ruegamer@stat.uni-muenchen.de}

\address[aaa]{Department of Statistics, LMU Munich, Ludwigstra{\ss}e 33, 80539, Munich, Germany}

\author[aaa]{Sonja Greven}

\begin{abstract}
Statistical inference after model selection requires an inference framework that takes the selection into account in order to be valid. Following recent work on selective inference, we derive analytical expressions for inference after likelihood- or test-based model selection for linear models.
\end{abstract}


\end{frontmatter}

\section{Introduction}

%
The invalidity of standard inference after model selection has been mentioned by many authors throughout the last decades, including \citet{Buehler.1963} and \citet{Leeb.2005}. Following these publications different approaches for inference in (high-dimensional) regression models after some sort of model selection have emerged over the past years. 
Initiated by the proposal for valid statistical inference after arbitrary selection procedures by \citet{Berk.2013}, many new findings and adoptions of post-selection inference (PoSI) to existing statistical methods have been published. Particularly notable is the general framework of \citet{Fithian.2014} transferring the classical theory of \citet{Lehmann.1955} in exponential family models to \emph{selective inference}. This post-selection inference concept is based on the conditional distribution of parameter estimators, conditional on the given selection event. Apart from general theory, several authors derive explicit representations of the space to which inference is restricted by well-known selection methods. Initially motivated by the application to the Lasso \citep[see, e.g.,][]{
Lee.2016} several recent publications aim for valid selective inference in forward stepwise regression or any forward stagewise algorithms. In this context, substantial work was done by \citet{Tibshirani.2014} as well as by \citet{Loftus.2014, Loftus.2015a} for linear models with known error variance $\sigma^2$. \citet{Tibshirani.2014} build a framework for any sequential regression technique resulting in a limitation to the space for inference, where the limitation can be characterized by a polyhedral set. \citet{Loftus.2014, Loftus.2015a} extend the idea to a more general framework, for which the limitation of the inference space is given by quadratic inequalities, which coincides with the polyhedral approach in special cases. 

Despite the popularity of the Lasso and similar selection techniques 
in statistical applications, likelihood-based model selection such as stepwise Akaike Information Criterion \citep[AIC,][]{Akaike.1973} selection is still used in an extremely vast number of statistical applications and diverse scientific fields \citep[see, e.g.,][]{
Zhang.2016
}. However, authors usually do not adjust their inference for model selection, although consequences may be grave \citep[see, e.g.,][]{Mundry.2009}. 
Selective inference allows to adjust inference after model selection, but an explicit representation of the required conditional distribution for likelihood-based model selection or similar selection procedures has not been derived so far.
%

We close this gap by explicitely deriving the necessary distribution in linear models with unknown $\sigma^2$ after likelihood- or test-based model selection, which comprises (iterative) model selection based on the AIC or Baysian Information Criterion \citep[BIC,][]{Schwarz.1978}, model selection via likelihood-based tests, F-tests, and p-value selection \citep[``significance hunting'',][]{Berk.2013} based on t-tests. We derive an analytical solution for inference in linear models after these model selection procedures and make available an \textsf{R} package for selective inference in such settings in practice \citep{coinflibs}. 
In addition, we provide inference for multiple and arbitrarily combined selection events, such as stepwise AIC selection followed by significance hunting.  We thereby close an important gap in the application of selective inference to model selection approaches that are ubiquitous in statistical applications across all scientific areas.
%

Section~\ref{sec:limo} presents the theory on selective testing for linear models and explicitely derives the necessary conditional distributions for several commonly used model selection approaches. 
In Section~\ref{sec:sim} we present simulation results for the proposed methods and apply our method to the prostate cancer data. 
We summarize our concept in Section~\ref{sec:disc}. Derivations of our results and visualizations of additional simulation settings can be found in the supplementary material online.

\section{Selective inference in linear models} \label{sec:limo}

After outlining the model framework and existing theoretical foundations on selective tests for linear models in Section~\ref{setup}, we present the new results on selective tests after various particular selection techniques in Section~\ref{expract} -- \ref{multsel}. We further show how to extend existing theory for the construction of conditional confidence intervals in Section~\ref{cis} and outline tests of grouped variables in this framework in Section~\ref{groups}.

\subsection{Setup and theoretical foundation} \label{setup}

Given $n$ independent variables $\bm{Y} = (Y_1,\ldots,Y_n)^\top$ with true underlying distribution $\mathcal{N}_n(\bm{\mu}, \sigma^2 \bm{I}_n)$, we consider as possible models submodels of the maximal linear model $$\bm{Y} = \bm{X} \bm{\beta} + \bm{\varepsilon}, \quad \bm{\varepsilon} \sim \mathcal{N}_n(\bm{0}, \sigma^2 \bm{I}_n)$$ for the given data $(\bm{y},\bm{X})$, where $\bm{y} = (y_1,\ldots,y_n)^\top$ are the observed values of $\bm{Y}$ and $\bm{X} = (\bm{x}_1, \ldots, \bm{x}_p) \in \mathbb{R}^{n \times p}$ is a fixed design matrix. In particular, we allow the considered models to be misspecified if $\bm{\mu}$ does not lie in the column space of the design matrix, in which case the corresponding model aims at estimating the linear projection of $\bm{\mu}$ onto the column space of the design matrix. We then compare two or more linear models based on different column subsets $\bm{X}_{\mathcal{T}}$ of $\bm{X}$ by using a likelihood-based model selection criterion, as for example the AIC. For the compared subsets, we let $\mathcal{T} \in \mathcal{P}(\{1, \ldots, p\})\backslash\emptyset$ with power set function $\mathcal{P}(\cdot)$. After selection of the ``best fitting'' model with design matrix $\bm{X}_{\mathcal{T}^\ast}$ with $|\mathcal{T}^\ast|=p_{\mathcal{T}^\ast}$, we would ideally like to test the $j$th regression coefficient in the set of corresponding coefficients $\bm{\beta}_{\mathcal{T}^\ast}$, i.e.
\begin{equation}
H_0: \beta_{\mathcal{T}^\ast \! ,j} = \theta. \label{eq:h0truth}
\end{equation}
However, taking into account that the true mean $\bm{\mu}$ is potentially non-linear in the selected covariates or the selection is not correct, we instead test the $j$th component of the projection of $\bm{\mu}$ into the linear space spanned by the selected covariates $\bm{X}_{\mathcal{T}^\ast}$:
\begin{equation}
H_0: \tilde{\beta}_{\mathcal{T}^\ast \! ,j} = \bm{v}^\top \bm{\mu} := \bm{e}_j^\top ({\bm{X}_{\mathcal{T}^\ast}}^\top \bm{X}_{\mathcal{T}^\ast})^{-1} {\bm{X}_{\mathcal{T}^\ast}}^\top \bm{\mu} = \theta, \label{eq:h0sel}
\end{equation}
where $\bm{e}_j$ is the $j$th unit vector and $\bm{v}$ is the so-called test vector. This coincides with (\ref{eq:h0truth}) if we select the correct model and $\bm{\mu}$ is actually linear in $\bm{X}_{\mathcal{T}^\ast}$. Testing the linear approximation instead of (\ref{eq:h0truth}) is a more realistic scenario in practice and is in line with the approach of several recent publications including \citet{Berk.2013}. 
%

We consider the following quadratic inequality introduced in a similar form by \citet{Loftus.2015a}, on the basis of which a model is chosen:
\begin{equation}
\bm{Y}^\top \bm{A} \bm{Y} + c \geq 0, \label{eq1}
\end{equation}
before showing that several common model selection approaches lead to restrictions on $\bm{Y}$ that can be written in this form. In most practical situations $c \equiv 0$. We are interested in the null distribution of $\hat{\beta}_{\mathcal{T}^\ast \! ,j} = \bm{v}^\top \bm{Y}$, which we use as a test statistic to test the null hypothesis (\ref{eq:h0sel}). 
Since $\bm{Y} \sim \mathcal{N}_n(\bm{\mu},\sigma^2 \bm{I}_n)$, $\bm{v}^\top \bm{Y} \sim \mathcal{N}_1(\bm{v}^\top \bm{\mu}, \sigma^2 \bm{v}^\top \bm{v})$ with $\bm{v}^\top \bm{\mu} = \theta$ under $H_0$. After model selection of the form (\ref{eq1}), $\bm{v}^\top \bm{Y}$ conditional on $\bm{Y}^\top \bm{A} \bm{Y} + c \geq 0$, and also conditional on $\bm{P}^\bot_{\bm{v}} \bm{Y} = \bm{P}^\bot_{\bm{v}} \bm{y}$ with $\bm{P}^\bot_{\bm{v}} \bm{y}$ the projection of $\bm{y}$ into the space orthogonal to $\bm{v}$, follows a truncated normal distribution \citep{Loftus.2015a} with truncation limits based on 
$\tau_{1/2} = \frac{1}{2}\delta^{-1}(-\zeta \pm \sqrt{\zeta^2 - 4\delta\xi})$, where $\delta = \bm{y}^\top \bm{P}_v \bm{A} \bm{P}_v \bm{y}$, $\zeta = 2 \bm{y}^\top \bm{P}_v \bm{A} \bm{P}_v^\bot \bm{y}$ and $\xi = \bm{y}^\top \bm{P}_v^\bot \bm{A} \bm{P}_v^\bot \bm{y} + c$. 
Due to the form of (\ref{eq1}), the two solutions $\tau_1 \leq \tau_2$ imply that the distribution of our test statistic is truncated to $(-\infty,  \tau_1 \cdot \bm{v}^\top \bm{y} ] \cup [ \tau_2 \cdot \bm{v}^\top \bm{y}, \infty)$ in the case in which $\delta$ is positive, and to $[\tau_1 \cdot \bm{v}^\top \bm{y}, \tau_2 \cdot \bm{v}^\top \bm{y}]$ if $\delta$ is negative. 


\subsection{Explicit derivations} \label{expract}

We now show that several commonly used model selection approaches can be written as in (\ref{eq1}) and explicitely derive the corresponding truncation limits to the normal distribution of the test statistic. For all derivations, please see the supplementary material. We always consider the comparison of two models $1$ and $2$ in which model $1$ is preferred over model $2$. Let $\mathcal{T}_k, k =1,2$ be the corresponding covariate subsets of the two considered models $k$ and let $\bm{X}_{k} := \bm{X}_{\mathcal{T}_k}$ denote the corresponding design matrix.

\textbf{Model selection based on log-likelihood comparison plus optional penalty term}. 
We start with conventional model selection procedures that are based on a log-likelihood comparison plus optional penalty term (as, for example, used in the AIC or BIC). Let $\ell_k$ be the log-likelihood of model $k$ and $\text{pen}_k$ the penalty term for this model, which is assumed not dependent on $\bm{Y}$. For example, if $p_k$ denotes the number of regression coefficients for model $k$ and the unknown $\sigma^2$ is estimated, $\text{pen}_k = 2(p_k + 1)$ for the AIC and $\text{pen}_k = \log(n) (p_k + 1)$ for the BIC. Furthermore, let $\hat{\sigma}_k^2$ be the scale parameter estimator and $\hat{\bm{\mu}}_k = \bm{P}_{X_k} \bm{Y}$ the mean vector estimator of model $k = 1,2$ with $\bm{P}_{X_k} = \bm{X}_k(\bm{X}_k^\top \bm{X}_k)^{-1} \bm{X}_k^\top$ and $\bm{X}_k \in \mathbb{R}^{n\times p_k}$. Then the model $1$ is selected iff
\begin{equation}
\begin{split}
& \quad -2 \ell_1(\bm{Y}) + \text{pen}_1  \leq -2 \ell_2(\bm{Y}) + \text{pen}_2 \\
\Leftrightarrow \quad & \quad \bm{Y}^\top \{ (n-p_1)\exp(-\gamma/n) (\bm{I}-\bm{P}_{X_2}) - (n-p_2) (\bm{I}-\bm{P}_{X_1}) \} \bm{Y} \geq 0 \label{IneqAIC2}
\end{split}
\end{equation}
with $\gamma = (p_2 - p_1 + \text{pen}_1 - \text{pen}_2)$. We therefore define $\bm{A} :=\{ (n-p_1)\exp(-\gamma/n) (\bm{I}-\bm{P}_{X_2}) - (n-p_2) (\bm{I}-\bm{P}_{X_1}) \}$ as well as $c := 0$. In the supplementary material we additionally derive the matrix $\bm{A}$ and $c$ when treating $\sigma^2$ as known and plugging in $\hat{\sigma}_1$, $\hat{\sigma}_2$ as estimators, i.e. when ignoring the fact that $\hat{\sigma}^2_k, k=1,2$ are also functions of $\bm{Y}$, to show the difference.

\textbf{Model selection on the basis of tests}.
We first consider the likelihood-ratio test (LRT). For model $1$ being nested in model $2$, the derivation is analogous to the AIC comparison by defining $\text{pen}_2 - \text{pen}_1 := q_{\chi^2_{1-\alpha} (p_2-p_1)}$, where $q_{\chi^2_{\alpha} (df)}$ is the $\alpha$-quantile of the $\chi^2$-distribution with $df$ degrees of freedom. 

The F-Test is not strictly likelihood-based, but falls into the same framework. Let $\text{RSS}_k = ||\bm{Y} - \hat{\bm{\mu}}_k||^2$ be the residual sum of squares of model $k$. If we choose model $1$, which is nested in model $2$, and denote by $F(\phi_1,\phi_2)$ the critical value of the F-distribution with $\phi_1$ and $\phi_2$ degrees of freedom, then:
\begin{equation}
\begin{split}
& \quad \frac{\frac{\text{RSS}_1 - \text{RSS}_2}{p_2 - p_1}}{ \frac{\text{RSS}_2}{n-p_2}} \leq F(p_2 - p_1, n-p_2)\\
\Leftrightarrow \quad & \quad \bm{Y}^\top \{ \bm{P}_{X_1} + \kappa (\bm{I}-\bm{P}_{X_2}) - \bm{P}_{X_2} \} \bm{Y} \geq 0, \label{IneqRSS}
\end{split}
\end{equation}
where $\kappa = F(p_2 - p_1, n-p_2) \cdot \frac{p_2 - p_1}{n - p_2}$ and therefore $\bm{A} = \{ \bm{P}_{X_1} + \kappa (\bm{I}-\bm{P}_{X_2}) - \bm{P}_{X_2} \}$ and $c = 0$. 
Similarly, if we select the larger model $2$ for either LRT or F-test, we simply have to invert the previous inequalities and define $\bm{A}$ as the negative of the respective matrices $\bm{A}$ defined above.

\textbf{``Significance hunting''}. 
As described in \citet{Berk.2013}, variable deselection or backward selection on the basis of the size of t-test p-values reduces to deselecting the smallest t-value among several candidates. For the comparison of two variables $j^\ast$ and $j$ and deselection of $j^\ast$ in the model $k$, it therefore holds that
\begin{equation}
|t_{j^\ast}| := \frac{|\hat{\beta}_{k,j^\ast}|}{\text{se}(\hat{\beta}_{k,j^\ast})} = \left| \frac{\bm{v}_{j^\ast}^\top \bm{Y}}{\sqrt{\hat{\sigma}_k^2 \bm{v}_{j^\ast}^\top \bm{v}_{j^\ast}}} \right| \leq \left| \frac{\bm{v}_{j}^\top \bm{Y}}{\sqrt{\hat{\sigma}_k^2 \bm{v}_{j}^\top \bm{v}_{j}}} \right|, \label{sighunt}
\end{equation}
where $\bm{v}_j^\top = \bm{e}_j^\top ({\bm{X}_k}^\top \bm{X}_k)^{-1} {\bm{X}_k}^\top$ and $\bm{v}_{j^\ast}^\top = \bm{e}_{j^\ast}^\top ({\bm{X}_k}^\top \bm{X}_k)^{-1} {\bm{X}_k}^\top$. Let $\bm{P}_{\bm{v}} = \bm{v}\bm{v}^\top / \sqrt{||\bm{v}||^2}$ for a given vector $\bm{v}$. Then (\ref{sighunt}) is equivalent to  
$\bm{Y} \left( \bm{P}_{\bm{v}_{j}} - \bm{P}_{\bm{v}_{j^\ast}}  \right) \bm{Y} \geq 0$ 
and we can define $\bm{A} := \left( \bm{P}_{\bm{v}_{j}} - \bm{P}_{\bm{v}_{j^\ast}}  \right)$, $c=0$. If only variables which are not significant are dropped for the ``significance hunting'', the t-value of $j^\ast$ additionally fulfills the condition $|t_{j^\ast}| \leq \mathcal{Q}_{T_{n-p_k}}(1-\frac{\alpha}{2})$, where $\mathcal{Q}_{T_{n-p_k}}(\cdot)$ is the quantile function of the Student's t-distribution with $n-p_k$ degrees of freedom, which is evaluated with a prespecified significance level $\alpha$ to obtain the decision. Since this is equivalent to $\bm{Y}^\top \bm{P}_{\bm{v}_{j^\ast}} \bm{Y} \leq \hat{\sigma}_k^2 \cdot (\mathcal{Q}_{T_{n-p_k}}(1-\frac{\alpha}{2}))^2$, we get $\bm{A} = \{ (\mathcal{Q}_{T_{n-p_k}}(1-\frac{\alpha}{2}))^2 (n-p_k)^{-1} (\bm{I}-\bm{P}_{\bm{X}_k}) - \bm{P}_{\bm{v}_{j^\ast}} \}$ and $c = 0$.
 
\subsection{Multiple selection events and p-value calculation} \label{multsel}

If there are $m$ selection events of the kind as in Section~\ref{expract}, the final space restriction can be calculated by finding the two (or more) most restrictive values in all limiting selection steps. Since this may involve several inequalities with different directions and may result in two or more non-overlapping intervals, additional care is needed. 
In general, let the resulting truncated normal distribution have multiple truncations given by the ordered intervals $[a_1, b_1], \ldots, [a_z, b_z], z \in \mathbb{N}$, where the case of no finite lower or upper truncation is given by $a_1 = - \infty$ or $b_z = \infty$ with intervals $(-\infty,b_1]$ or $[a_z,\infty)$ implied by convention, respectively. Let $\hat{\beta}_{\mathcal{T}^\ast \! ,j} = \bm{v}^\top \bm{y}$ be the observed value of the test statistic, which lies in the interval $[a_l, b_l]$ for some $l \in \{1,\ldots,z \}$. Then, following \citet{Tibshirani.2015}, a p-value $p\sim\mathcal{U}[0,1]$ for the two-sided significance test for (\ref{eq:h0sel}) based on $\hat{\beta}_{\mathcal{T}^\ast \! ,j}$ can be calculated via 
$p = 2 \cdot \min(\tilde{p}, 1-\tilde{p})$, 
with $\tilde{p}$ being the p-value of the one sided test. In our setting and as we allow for multiple disjoint truncation intervals, we can define this as 
$\tilde{p} = \mathbb{P}_{H_0}(\bm{v}^\top \bm{Y} > \hat{\beta}_{\mathcal{T}^\ast \! ,j} \mid \text{selection event}, \bm{P}_v^\bot \bm{Y} = \bm{P}_v^\bot \bm{y} ) = \Psi_{\text{nom}} / \Psi_{\text{denom}}$, 
where 
$\Psi_\text{nom} = \psi(b_l) -\psi(\hat{\beta}_{\mathcal{T}^\ast \! ,j}) + \sum_{i = l+1}^{z}\, \psi(b_i) - \psi(a_i)$,
$\Psi_\text{denom} = \sum_{i = 1}^z \psi(b_i) - \psi(a_i)$ and
$\psi(x) = \Phi(\frac{x}{\sigma\sqrt{\bm{v}^\top \bm{v}}})$ with cumulative distribution function $\Phi(\cdot)$ of the standard normal distribution. 
In other words, $\Psi_\text{denom}$ is equal to the cumulative probability mass for all possible values $\bm{v}^\top \bm{Y}$ that comply with the conditioning event, and $\Psi_\text{nom}$ is the cumulative probability mass of possible values $\bm{v}^\top \bm{Y}$ that are larger than $\hat{\beta}_{\mathcal{T}^\ast \! ,j}$. 

As in practice $\sigma^2$ is usually unknown, we investigate in simulations the performance and validity of our proposed p-values when plugging in the restricted maximum likelihood estimate $\hat{\sigma}_\text{REML}^2 = ||\bm{y}-\bm{X}_{\mathcal{T}^\ast} \hat{\bm{\beta}}_{\mathcal{T}^\ast}||^2 / (n-p_{\mathcal{T}^\ast})$ for $\sigma^2$. We describe the corresponding results in Section~\ref{sec:sim}. Note that while $\hat{\sigma}_\text{REML}^2$ is plugged into the truncated normal conditional distribution for $\bm{v}^\top \bm{Y}$, this distribution is exact and does account for estimation of $\sigma^2$ in the selection event.

\subsection{Conditional confidence intervals} \label{cis}

We extend the results of \citet{Tibshirani.2014} to allow for the construction of selective confidence intervals if the null distribution is truncated to several intervals. We thus find the quantiles $q_{\alpha/2}$ and $q_{1-\alpha/2}$, for which $$\mathbb{P}(q_{\alpha/2} \leq \bm{v}^\top \bm{\mu} \leq q_{1-\alpha/2} \mid \text{selection event}, \bm{P}_v^\bot \bm{Y} = \bm{P}_v^\bot \bm{y}) = 1 - \alpha.$$ Analogous to \citet{Tibshirani.2014}, we can make use of the fact that the truncated normal survival function with multiple truncation limits is also monotonically decreasing in its mean $\theta$ as the truncated normal distribution with multiple truncation intervals is a natural exponential family in $\theta$ \citep[see][]{Fithian.2014, Lee.2016}. 
The corresponding quantiles can be found via a grid search, where $q_\alpha$ satisfies $1-F^{\cup_l [a_l,b_l]}_{\mathcal{N}(q_\alpha,\sigma^2 \bm{v}^\top \bm{v})}(\bm{v}^\top \bm{y}) = \alpha$ with $F^{\mathcal{J}}_{\mathcal{N}(\mu,\sigma^2)}$ being the truncated cumulative normal distribution function with mean $\mu$, variance $\sigma^2$ and truncation interval(s) $\mathcal{J} \subseteq (-\infty, \infty)$. In other words, we search for the mean values $\theta = q_{\alpha/2}$ and $\theta = q_{1-\alpha/2}$ of the truncated normal distribution $\mathcal{N}^{\mathcal{J}}(\theta, \sigma^2 \bm{v}^\top \bm{v})$, for which the observed value $\bm{v}^\top \bm{y}$ is equal to the $\alpha/2$ and $1-\alpha/2$ quantile, respectively, and $H_0: \beta_{\mathcal{T}^\ast \!,j} = \theta$ thus would not be rejected. 

\subsection{Testing groups of variables} \label{groups}

Following \citet{Loftus.2015a}, a selective $\chi$-significance test for groups of variables can be constructed by testing the null hypothesis $H_0: \tilde{\bm{P}}_g \bm{\mu} = \bm{0},$ where 
$\displaystyle\tilde{\bm{P}}_{g} = \tilde{\bm{X}}_{\mathcal{T}^\ast\!,g} \, (\tilde{\bm{X}}_{\mathcal{T}^\ast\!,g}^\top \, \tilde{\bm{X}}_{\mathcal{T}^\ast\!,g})^{-1}\, \tilde{\bm{X}}_{\mathcal{T}^\ast\!,g}^\top,$ 
$\tilde{\bm{X}}_{\mathcal{T}^\ast\!,g} = (\bm{I} - \bm{P}_{\mathcal{T}^\ast \backslash g}) \bm{X}_{\mathcal{T}^\ast\!,g}$,
$\bm{X}_{\mathcal{T}^\ast,g}$ are the columns of the grouped variable $g$ in $\bm{X}_{\mathcal{T}^\ast}$, 
$\bm{P}_{\mathcal{T}^\ast \backslash g}$ is the projection onto the column space of $\bm{X}_{\mathcal{T}^\ast \backslash g}$ and 
$\bm{X}_{\mathcal{T}^\ast \backslash g}$ are the columns of $\bm{X}_{\mathcal{T}^\ast}$ without $\bm{X}_{\mathcal{T}^\ast\!,g}$. 
Without model selection, a test statistic is given by $T = \sigma^{-1} ||\tilde{\bm{P}}_g^\top \bm{Y}||_2 \overset{H_0}{\sim} \chi_{\text{Trace}(\tilde{\bm{P}}_g)},$ i.e., $T^2$ follows a $\chi^2$-distribution with $\text{Trace}(\tilde{\bm{P}}_g)$ degrees of freedom under $H_0$. When conditioning on $(\bm{I} - \tilde{\bm{P}}_g) \bm{Y} = (\bm{I} - \tilde{\bm{P}}_g) \bm{y} =: \bm{z}$ and the unit vector $\bm{u}$ in the direction of $\tilde{\bm{P}}_g^\top \bm{y}$, $\bm{Y}$ can be decomposed as $\bm{Y} = \bm{z} + \sigma T \bm{u}$, such that the only variation is in ${T}$. Conditional on the selection event (\ref{eq1}), $T$ follows a truncated $\chi$-distribution with truncation limits $\tau_{1/2}$ now given by $\delta = \sigma^2 \bm{u}^\top \bm{A} \bm{u}$, $\zeta = 2\sigma \bm{u}^\top \bm{A} \bm{z}$ and $\xi = \bm{z}^\top \bm{A} \bm{z} + c$. 
Depending on the sign of $\delta$ and the number of solutions $\tau_{1/2} \geq 0$, the truncation set $\mathcal{J} \subseteq [0,\infty)$ is either a closed interval $\mathcal{J} = [\max(0,\tau_1), \tau_2]$, an open interval $\mathcal{J} = [\tau_2, \infty)$, or a union of intervals $\mathcal{J} = [0,\tau_1] \cup [\tau_2,\infty)$.
The test for grouped variables with multiple selection events can be treated analogously to Section~\ref{multsel} by normalizing the truncated $\chi$ distribution analogously, replacing $\psi$ with the cumulative distribution function of the $\chi_{\text{Trace}(\tilde{\bm{P}}_g)}$-distribution. Note that while the truncated normal distribution is replaced by a truncated $\chi$-distribution, the types of conditioning events do not change when incorporating groups of variables. The only exception is \emph{significance hunting}, for which model selection is then not based on t-statistics of regression coefficients but an F-test as in (\ref{IneqRSS}) is typically used. 

\section{Empirical evidence} \label{sec:sim}

We evaluate the proposed selective inference concepts in linear models for a forward stepwise selection procedure based on the AIC.

\begin{figure}[ht]
\centering
\includegraphics[width=\linewidth]{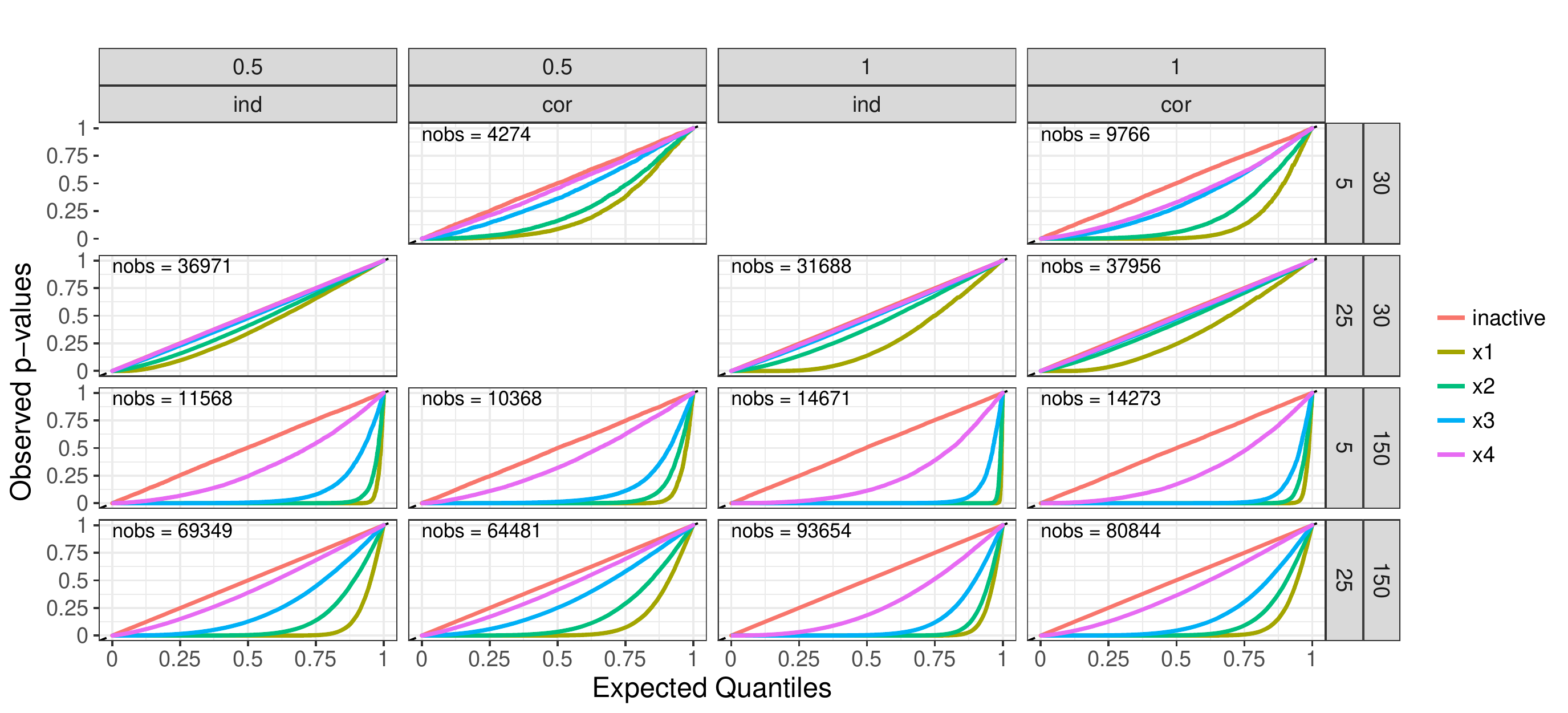}
\caption{Quantiles of the standard uniform distribution versus the observed p-values for different SNR and correlation settings (columns) as well as different settings for n and p (rows) in simulation iterations in which all of the active covariates and additional inactive covariates are selected. Combinations with no visualization indicate that no iterations are available for this simulation setting in which this holds. p-values were calculated on the basis of the true variance. For each setting, the number of iterations (nobs) is noted in the left upper corner.}\label{fig:plot}
\end{figure}

For the simulation study, we consider $p \in \{ 5, 25 \}$ covariates $\bm{x}_1, \ldots, \bm{x}_p \in \mathbb{R}^n$, $n \in \{30, 150 \}$ observations and use the data generating process $\bm{y} = \bm{X}^\dagger \bm{\beta}^\dagger + \bm{\varepsilon}.$ $\bm{X}^{\dagger} = (\bm{x}_1, \ldots, \bm{x}_4)$ respectively $\bm{\beta}^\dagger = (4,-2,1,-0.5)^\top$ correspond to the true active covariates respectively their effects and $\bm{\varepsilon}$ is Gaussian noise with zero mean and variance $\sigma^2$, which is determined by the signal-to-noise ratio $\text{SNR} \in \{0.5, 1\}$. Covariates are independently drawn from a standard normal distribution ($ind$) or exhibit a correlation of $0.4$ ($cor$). 
For each setting, $100,000$ simulation iterations are performed. We present resulting p-values in a \emph{uniform quantiles vs. observed p-value}-plot in Figure \ref{fig:plot}, where p-values are calculated on the basis of concepts introduced in Section~\ref{sec:limo}. In the plot, p-values along the diagonal indicate uniformity, which seems to hold for all inactive variables in all given simulation settings. For active variables, the corresponding selective test shows higher power the closer the point line of p-values runs along the axis. Results are based on those simulation iterations in which all of the active covariates and additional inactive covariates are selected. 
Note that in the selective inference framework, p-values of inactive variables should exhibit uniformity given any particular set of selection events, if the null hypothesis holds. Aggregating across selected models in each panel of Figure~\ref{fig:plot} results in mixture distributions for the p-values, with a mixture of uniform $\mathcal{U}[0,1]$ variables again being $\mathcal{U}[0,1]$.
Results for iterations without selected inactive variables (not shown) are similar in terms of power.
In summary, 
p-values for inactive variables exhibit uniformity in every setting. p-values for active covariates indicate large power in most of the settings, with notable exceptions for those simulation settings in which $p$ is relatively large in comparison to $n$.

Further results are given in the supplementary material, showing the resulting p-values for simulation iterations in which the selected model is 
misspecified due to missing active variables and potentially selected inactive variables. 
Here, p-values of inactive variables exhibit some deviation from the uniform quantiles when not all of the active variables have been selected. However, deviations mainly occur when inactive variables are correlated with unselected active variables in which case the null hypothesis (\ref{eq:h0sel}) in fact does not exactly hold. This is due to the fact that the linear projection of $\bm{\mu}$ into the column space of the selected design matrix has a non-zero coefficient for the $j$th variable if a correlated variable is omitted from the model.
For the setting with correlation, $n=150$, $p=25$ and $\text{SNR}=1$, Table~\ref{tab2} additionally provides the estimated coverage for the confidence intervals constructed as in subsection~\ref{cis}, averaging over all iterations where at least all the active variables are selected (and over inactive variables for the inactives column). 
In addition, we investige the performance of our approach when plugging in $\hat{\sigma}_\text{REML}^2$ for $\sigma^2$ in the derived distribution of $\hat{\beta}_{\mathcal{T}\! ,j}$ for all simulation settings (see supplementary material). p-values for inactive variables still approximately exhibit a uniform distribution when using an estimate for $\sigma^2$. Notable deviations in comparison to p-values calculated with the true variance can occur when $\sigma^2$ is not estimated well such as for $n=30$ and $p=25$. 
Furthermore, as shown in Table~\ref{tab2}, almost no difference in the coverage of selective confidence intervals is obtained when plugging in $\hat{\sigma}_\text{REML}^2$ for $\sigma^2$.
In the supplementary material, we also provide results for a simulation study for the $\chi$-test after stepwise AIC selection with a group noise variable. 

\begin{table}[ht]
\caption{Coverage of selective 95\% confidence intervals for the simulation setting with correlation, $n=150$, $p=25$ and $\text{SNR}=1$ for selection cases in which all the active (and potentially additional inactive) variables are selected after AIC stepwise forward selection. The coverage is estimated using $8725$ observations for active and $31371$ observations for inactive variables.} \label{tab2}
\centering
\begin{tabular}{rrrrrr}
& Inactives & $x_1$ & $x_2$ & $x_3$ & $x_4$ \\ 
  \hline
Using true variance & 0.9516 & 0.9492 & 0.9485 & 0.9532 & 0.9542 \\
Using plugin estimate & 0.9496 & 0.9485 & 0.9457 & 0.9515 & 0.9532\\
   \hline
\end{tabular}
\end{table}

We additionally apply our approach to the prostate cancer data set \citep{Stamey.1989}, which has also been used in \citet{Tibshirani.2014} to illustrate selective confidence intervals after forward stepwise regression (see the supplementary material). When using $\alpha = 0.05$, the significant variables match the two significant variables after forward stepwise regression in \citet{Tibshirani.2014}, although the selected model is different. Compared to unadjusted inference, confidence intervals become wider for all coefficients in the selective inference framework.

\section{Summary} \label{sec:disc}

Based on the general selective inference framwork derived in \citet{Loftus.2014, Tibshirani.2014, Loftus.2015a}, we address the issue of conducting valid inference in linear models after likelihood- or test-based model selection, which comprises (iterative) model selection based on the AIC or BIC, model selection via likelihood-based or F-tests and significance hunting based on t-tests. We explicitely derive the necessary conditional distributions for these selection events, which allows the application of selective inference to additional practically relevant settings compared to existing results. We extend the construction of p-values and confidence intervals to the case in which the distribution of the test statistic conditional on the selection events is truncated to multiple intervals. In simulations, we see that obtained p-values yield desireable properties 
even if the selected model is not correctly specified and confidence intervals show the nominal coverage. We make available an \textsf{R} software package \citep{coinflibs} for selective inference to apply the proposed framework in practice.


%
%
%
%
%

\section*{Acknowledgments}

We thank Fabian Scheipl for his useful comments.

\section*{References}

\bibliography{mybibliography}

\end{document}